\documentclass[superscriptaddress,amssymb,twocolumn,aps]{revtex4}

\usepackage{graphicx,dcolumn,bm,amsmath,amssymb,color,bm}

\usepackage[utf8]{inputenc}
\usepackage[T1]{fontenc}
\usepackage{mathptmx}
\usepackage{gensymb}
\usepackage{soul,xcolor}
\setstcolor{red}

\newcommand{\JDM}[1]{\textcolor{black}{#1}}

\usepackage{hyperref}
\usepackage{color}

\newcommand{\revv}[1]{ {\color{black} #1}}

\definecolor{amber}{rgb}{1.0, 0.75, 0.0}

\newcommand{\eeq}{ \end{equation} }
\newcommand{\beq}{ \begin{equation} }

\newcommand{\eea}{ \end{align} }
\newcommand{\bea}{ \begin{align} }

\usepackage{xr}
\makeatletter
\newcommand*{\addFileDependency}[1]{
  \typeout{(#1)}
  \@addtofilelist{#1}
  \IfFileExists{#1}{}{\typeout{No file #1.}}
}
\makeatother
\newcommand*{\myexternaldocument}[1]{%
    \externaldocument{#1}%
    \addFileDependency{#1.tex}%
    \addFileDependency{#1.aux}%
}
\myexternaldocument{supplementary}

\begin{document}

\preprint{AIP/123-QED}


\title{Melting of rods on a sphere via an intermediate hexatic phase}

\author{Jaydeep Mandal}

\author{Chandan Dasgupta}

\author{Prabal K. Maiti}
\email{maiti@iisc.ac.in}
\affiliation{%
Centre for Condensed Matter Theory, Department of Physics, Indian Institute of Science, Bengaluru 560012, India
}%
\date{\today}

\begin{abstract}

 We have studied, using molecular dynamics simulations, the pressure-induced melting in a monolayer of soft repulsive spherocylinders whose centers of mass are constrained to move on the surface of a sphere. We show that the orientational degrees of freedom of the spherocylinders exhibit nematic order, whereas the positions of their centers of mass exhibit melting transitions that depend on the radius of the confining spherical surface. \JDM{Our system presents a unique scenario where the decoupling of the orientational degrees of freedom from the positional degrees of freedom leads to an effectively two-dimensional (2D) crystal-to-liquid transition \JDM{on a spherical surface}. Further study of the nature of this \JDM{2D} melting on a sphere shows that the transition is a two-step process, and there exists a very small window of an intermediate hexatic phase between crystal and liquid phases. Similar results are found for flat monolayers (with the radius of the sphere \(R \rightarrow \infty\)). We show that, interestingly, the structure of the defects, originating from the curvature of the substrate, also changes during melting.}
 
\end{abstract}

\maketitle


\section{\label{sec:intro}Introduction:}

Phase transitions \cite{landau1937theory} are fascinating subjects studied in physics, and the nature of phase transitions can be different depending on various factors such as the symmetry of the order parameter and the dimension of space. In 1973, Kosterlitz and Thouless \cite{kosterlitz2018ordering,kosterlitz1978two,kosterlitz2016kosterlitz} showed that phase transitions in two-dimensional (2D) systems may be fundamentally different from those in three dimensions. Similar concepts were developed by Halperin and Nelson \cite{halperin1978theory,nelson1979dislocation} and Young \cite{young1979melting} in the context of 2D melting. In the proposed description, topological defects play a major role in the transitions. The ordered crystal transforms to a phase with free dislocations at a temperature higher than the melting temperature (\(T_m\)). When the temperature is increased further, the dislocations dissociate and disclination unbinding takes place. When the temperature is higher than a specific temperature \(T_i\), the free disclinations dominate and the system melts to an isotropic liquid. Hence, according to the proposed theory, the transitions in 2D occur via two steps, and the intermediate phase is often termed as "hexatic". Such phase transitions in 2D are called KTHNY (Kosterlitz-Thouless-Halperin-Nelson-Young) transitions. Note that Berezinskii \cite{berezinskii1971zh,berezinskii1972zh} is often considered to have proposed the basics of the theory before others, and sometimes the transition is named after him as well (BKTHNY transition).

The BKTHNY transitions seemed pretty general in their framework, and very soon the search for the hexatic phase began. Many theoretical, experimental and computer simulation studies were performed which claimed the existence of the hexatic phase or otherwise. Experimental systems of rare gas atoms adsorbed on a graphite surface \cite{brinkman1982melting,strandburg1988two,taub1977neutron,heiney1982freezing,specht1984phase,mctague1982synchrotron} claimed to show a continuous transition. which was later contradicted by computer simulations \cite{koch1983freezing,abraham1983melting,abraham1984melting}. Many other experimental systems with polystyrene spheres \cite{murray1987experimental,skjeltorp1983one,armstrong1989isothermal}, and others \cite{deutschlander2013two,horn2013fluctuations,gasser2010melting,zahn2000dynamic,keim2007frank,von2004elastic,dillmann2008polycrystalline,assoud2009ultrafast,deutschlander2014specific} showed agreement with BKTHNY theory. In computer simulations, contradictory results were obtained for different inter-particle potentials used in the calculations. Contrary to experiments, many simulation studies reported first-order transition \cite{kalia1981interfacial,zollweg1992melting,lee1992first,allen1983monte,novaco1982relaxation,abraham1980melting,barker1981phase,toxvaerd1980phase,weber1995melting,broughton1982molecular}, while some suggested two-step melting \cite{naidoo1994melting,frenkel1979evidence,zollweg1989size,chen1995melting,bagchi1996computer,bagchi1996observation}. Recent studies by Krauth et. al \cite{bernard2011two,kapfer2015two} remarkably suggested that for 2D hard disc systems, melting occurs via a continuous crystal-to-hexatic transition obeying the BKTHNY mechanism, while the hexatic-to-liquid transition is first order. The experimental evidence for this proposition came in 2017 \cite{thorneywork2017two}. The generally obeyed rule is that for long-range potentials, the transitions are BKTHNY type, whereas for short-range interactions, a scenario of continuous crystal-hexatic and first order hexatic-isotropic transitions can take place.

Although the existence of the hexatic phase in flat 2D surfaces is still debated but well-studied, a less-trodden ground is the study of the nature of phase transitions on curved surfaces. Curvatures play an important role in the phase behaviors of

various physical, biological and other systems \cite{keller2008reconstruction,collinson2002clonal,bates2008nematic,dhakal2012nematic,venkatareddy2023effect}. From the Euler theorem, triangulation is only possible on a spherical surface in the presence of defects, the total charge of which has to be \(12\). This is attainable with the existence of twelve "5-fold" defects at the vertices of an icosahedron, as observed in a football or virus capsid \cite{bruinsma2021physics}. Such defects can also show "domain scar" \cite{bausch2003grain}. A Very recent experimental study \cite{singh2022observation} shows the existence of the hexatic phase and the role of topological defects while melting from defect-ridden crystal to isotropic liquid. 

In the manuscript, we have investigated the pressure-induced melting transition of soft repulsive spherocylinders (SRS) forming a spherical monolayer. Very recently, a similar system was experimentally realized \cite{devries2007divalent}.  We report our main results below and present the simulation techniques used in this work in the "Materials and Methods" section. We demonstrate the existence of different phases for a spherical monolayer of spherocylinders and calculate the phase transition packing fractions. Our calculations are consistent with the results for a planar monolayer (\(R \rightarrow \infty\) with \(R >> L\)), where \(R\) is the radius of the spherical substrate and \(L\) is the length of the spherocylinders. We also show, using different analyses involving hexatic order parameter, structure factor and susceptibilities that there exists an intermediate hexatic phase for a very small packing fraction range (\(\Delta \eta = 0.02\)) during the melting transition in both spherical and flat monolayers when the system size is large. We also show that the structure of the defects, arising due to the spherical topology of the confining surface, also changes during the melting transition. At the end we summarize these results in section \ref{sec:conclusion}.

\section{Results} \label{sec:results}

\begin{figure*}
\centering
\includegraphics[width=1.0\linewidth]{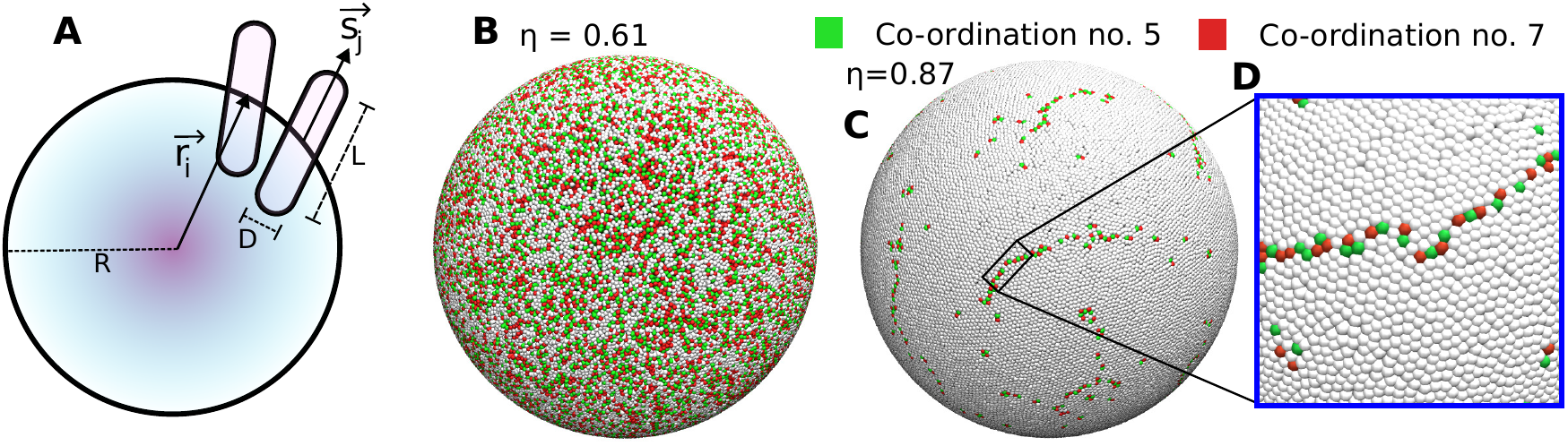}
\caption{Schematics of the studied system: A) The spherocylinders, with body-length \(L\), diameter \(D\) \JDM{(taken to be \(1.0\))}, and aspect ratio \(A = L/D\), have their centers of mass constrained on a sphere of radius \(R\). The long axis of the spherocylinders is along \(\Vec{s}\), the position of the center of masses is denoted by \(\Vec{r}\). B) and C) The equilibrium phases for the system and the corresponding snapshots of the system at different packing fractions are presented for \(N=25000, A=2.0, T^*=5.0\). A medium-density Liquid crystalline phase where the rods are aligned radially outwards, and the number of topological defects is high is shown in B. C) A high-density crystalline phase with positional and orientational ordering, with dislocations. D) Zoomed-in version of the defect structures: the packing is hexagonal, although a number of disclination defects with opposite charges form "domain scars". The green and red colors indicate the particles with co-ordination number of \(5\) and \(7\) respectively. The visualizations were obtained using the software VMD \cite{humphrey1996vmd}.}
\label{fig:snapshot}
\end{figure*}

 This work deals with the pressure-induced melting of soft repulsive spherocylinders in a spherical monolayer, where we have reduced the pressure of the system slowly such that the density of the system decreases at a fixed temperature. The system can be described by: (see fig \ref{fig:snapshot} A) a) the 3D orientation vector of the spherocylinders denoting their long axis, \(\Vec{s_i} = (s_{i1},s_{i2},s_{i3})\), where \(i\) is the index of a spherocylinder and b) the 2D position of the center of masses of the spherocylinders, \(\Vec{r_i} = (x_i,y_i,z_i)\), where \(x_i^2 + y_i^2 + z_i^2 = R^2\), and \(R\) is the radius of the sphere. Because of this constraint equation, the position vectors are effectively two-dimensional. For the case of flat monolayer, \(R \rightarrow \infty\), \(\Vec{r_i}\) is purely 2D, whereas \(\Vec{s_i}\) remains 3D for all cases. The vectors are defined with respect to a co-ordinate system, the origin of which is located at the center of the sphere. \JDM{The spherocylinders have body-length \(L\), and diameter \(D\) (fig \ref{fig:snapshot}A), where \(D = 1.0\) is considered for the convenience of calculation. The aspect ratio of the spherocylinders is given by \(A = L/D = L\).} The packing fraction of the system is defined as \(\eta = \frac{N}{16R^2}\), where \(N\) is the total number of spherocylinders, \JDM{which was kept fixed for a particular set of simulations}.
 
 In this section, we discuss the main results obtained for the pressure-induced melting of rods on a spherical monolayer. We first identify the phases and phase transition densities and study the effect of curvature-induced strain on the transition densities. When we decouple the orientation degrees of freedom of the spherocylinders which span a third dimension and study the phase transition from the reference frame of the centers of masses (which lie on a sphere), we find a crystal-to-liquid melting transition. We elaborate on whether an intermediate hexatic phase can be observed for this (pseudo)2D melting transition. We provide evidence supporting that there is an intermediate hexatic phase for such pressure-induced melting of rods on a spherical, as well as for a flat monolayer for which \(R \rightarrow \infty\). 

 \subsection{Phases} \label{subsec:phases}

  As has been demonstrated by our earlier work \cite{rajendra2023packing}, a spherical monolayer of SRS particles exhibits two phases (fig \ref{fig:snapshot}B,C)
  - a) a Liquid Crystal (LC) phase at low to medium packing fractions. In this phase, we observe no positional ordering of the centers of mass of the particles in the plane of the sphere, but there exists a finite out-of-plane orientational order. b) \JDM{An ordered} phase where both in-plane positional and out-of-plane orientational order exist. \JDM{The true nature of the ordered phase can be crystalline or hexatic from the reference frame of the sphere, the discussion of which we will postpone to the section \ref{sec:hexatic_intro} of this manuscript, and for convenience, we will denote this ordered phase as crystalline or "K". } 

  We note that for such cases of confined systems, the interesting phenomenon of the non-existence of the isotropic phase even at low densities is observed, which occurs due to the effect of confinement. The non-existence of isotropic phase can be verified by the high value of the Nematic order parameter, \(S_{nem}\) (an order parameter to quantify the orientational order in the system), at low densities and also from the fact that the nematic order parameter does not show any scaling relation (namely $S_{nem} \propto 1/\sqrt{N}$) with the number of particles in the system, even at low packing fractions where the isotropic phase typically exists for bulk 3D cases \cite{bolhuis1997tracing,chattopadhyay2021heating}. The physical reason behind this is entropic, which prefers the out-of-plane ordered structure of the system to maximize the available free surface per particle and thus the entropy.

  Another interesting effect observed due to the spherical confinement is the presence of topological defects in the system. For the systems that we have worked with, the defects are disclinations which consist of particles having a coordination number of \(5\) ("+1 defect") and \(7\) ("-1 defect")(\cite{bowick2009two}) (compared to the hexagonal packing's "pure" co-ordination number \(6\)). The two nearest-neighbour disclinations with opposite charges form a dislocation. We observe that for high density, when the \JDM{ordered} phase exists, the dislocations are attracted to each other and form a "domain boundary line" (see fig \ref{fig:snapshot}D) on the spherical surface. On the other hand, the liquid crystalline phase configurations show no in-plane positional order, which indicates that the centers of mass of the particles exhibit a larger number of randomly distributed defect points (fig \ref{fig:snapshot}B).

\subsection{Phase Transition and Transition Packing Fraction}

\begin{figure}
\centering
\includegraphics[width=1.0\linewidth]{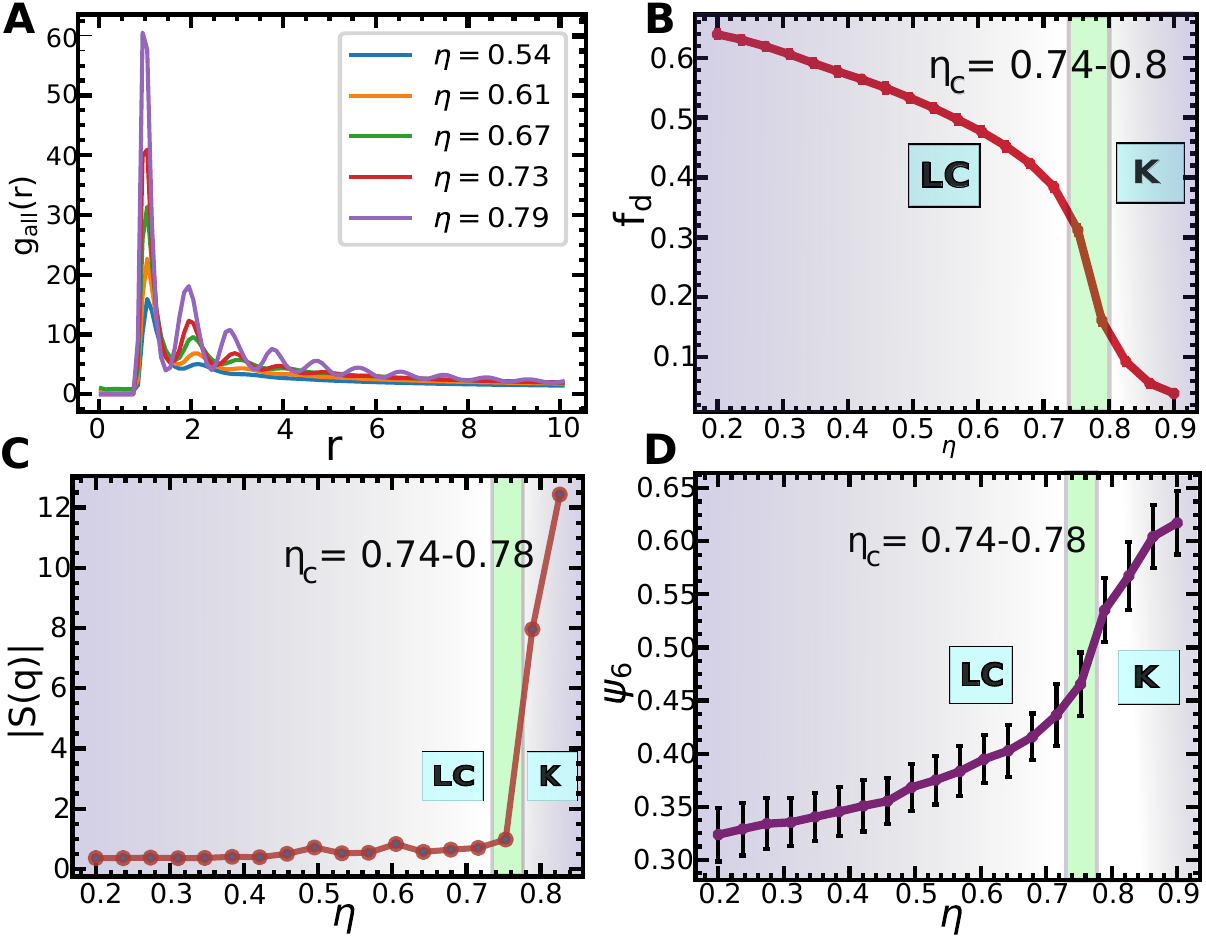}
\caption{Calculation of the transition packing fraction from Liquid Crystal (LC) to \JDM{ordered or} crystal (K) phase for \(N=25000,A=5.0\): A) the plot of radial distribution function of the centers of mass of the system on the sphere, as a function of euclidean distance. The gradual emergence of more number of peaks at high densities indicates the crystal phase. 
B) The total number of defects in the system decreases with packing fraction, as the crystal phase is approached. The green-shaded region is used to indicate the transition packing fraction range. In this region, the total number of defects changes sharply. C) The intensity of the structure factor plotted as a function of packing fraction. The sharp change in value around \(\eta_c 
 \approx 0.74-0.78\) indicates a transition to \JDM{ordered} crystal phase. D) The hexatic order parameter as a function of packing fraction. The transition density indicated by this calculation is around \(\eta_c \approx 0.74-0.78\)}
\label{fig:etac_calculate}
\end{figure}

We next try to evaluate the packing fraction at which the transition from the LC phase to the (K) phase takes place. We denote the transition packing fraction by \(\eta_c\). 
\subsubsection{Radial Distribution function (RDF)}
To precisely obtain the values of \(\eta_c\), we first computed the radial distribution function (RDF) of the centers of mass of the particles (fig \ref{fig:etac_calculate}A). The LC and K phases differ via the existence of positional order in the plane of the sphere. Hence the RDF shows the gradual emergence of a number of peaks with distance, indicating the appearance of long-range positional order, as the packing fraction is increased. From Fig \ref{fig:etac_calculate}A we observe that for \(N=25000, A=5.0\) at a packing fraction \(\eta = 0.73\), liquid-like behavior is observed. As the packing fraction is increased, the RDF shows repeated peaks characteristic of an \JDM{ordered} phase at \(\eta=0.79\). Hence we determine the transition packing fraction \(\eta_c\) for \(N=25000,A=5.0\) to lie in the range \(0.73-0.79\)
\subsubsection{Defect count}
In order to determine the transition packing fraction for \(N = 25000, A = 5.0\), we have also used the fact that the number of defect points in the LC phase drastically increases from the \JDM{ordered} phase (as evident from fig \ref{fig:snapshot}B and C). Thus, we have calculated the total number of defect points defined as 

\begin{equation}
 f_d =1 - \frac{N_6}{N}
 \label{eqn:total_fd} \\
\end{equation}

where \(N_6\) is the total number of particles which have the co-ordination number \(6\). In fig \ref{fig:etac_calculate}B), we show that as the packing fraction for a particular system increases, the total number of defect points decreases. We observe a sharp change in the number of defects at a packing fraction \(\eta\) in a range of \(0.74-0.80\). This sharp change indicates the transition of the system from the LC to the K phase and the range of \(\eta\) value matches with the estimated range from the RDF analysis. 

\subsubsection{Intensity of structure factor}
To confirm the identification of the transition region, we also calculated the static structure factor for the system for different packing fractions. The structure factor of the system is computed using the following relation:

\begin{equation}
    S(\Vec{q}) = \frac{1}{N} \langle \sum_{i=1}^N\sum_{j=1}^{N} e^{-i\Vec{q} \cdot \left(\Vec{r}_i -\Vec{r}_j\right)} \rangle  \label{eqn:str_fac} \\    
\end{equation}

 where \(\Vec{q}\) is a reciprocal lattice vector and \(\Vec{r_i}\) and \(\Vec{r_j}\) are the center-of-mass position vectors of \(i-th\) and \(j-th\) spherocylinders.

 The structure factor shows a liquid-like homogeneous behavior in the reciprocal space at low packing fractions and sharp Bragg-like peaks begin to appear at higher packing fractions. The intensity of the first peak (or the height of the first peak), also increases with the packing fraction. The intensity of the first peak is plotted as a function of packing fraction in fig \ref{fig:etac_calculate}C. The sharp rise in the value of the intensity indicates the transition from a Liquid-Crystal (LC) to a \JDM{ordered} (K) phase. The transition, as indicated in fig \ref{fig:etac_calculate}C), occurs around \(\eta_c \approx 0.74-0.78\).

\subsubsection{Bond-orientational order parameter}
We also corroborated the earlier stated transition packing fraction values (\(\eta_c\)) with the calculation of the bond-orientational order parameter or hexatic order parameter \(\Psi_6\) for the center-of-mass positions of the particles on the sphere. The bond-orientational order parameter is defined as \cite{jami2024effect}:
\begin{equation}
\Psi_6 =\frac{1}{N}\Sigma_j \psi_j \quad
\psi_j =\frac{1}{n}\Sigma_{k=1}^n exp(i 6\phi_{jk})  \label{eqn:psi6} \\
\end{equation}
where \(n\) is the number of nearest neighbor of the \(j-th\) particle. \(\phi_{jk}\) is the angle between the bond vector joining the \(j-th\) particle with its \(k-th\) neighbor and a specific axis.

Note that due to the curvature of the system, we have calculated the hexatic order parameters of several subsystems on different faces on the sphere and taken average over the faces for better statistics, and hence the specific axis will change depending on the face of the sphere. Also, for clarity, we state that the bond-orientational order parameter, which measures the orientational order of the centers of mass of the particles in 2 dimensions, is different from the nematic order parameter \(S_{nem}\), which measures the orientational order of the long axes of the spherocylinders in 3 dimensions. In subsequent sections, the terms "orientational order" and "bond-orientational order" are used interchangeably and indicate the 2D bond-orientational order.

The bond-orientational order parameter \(\psi_6\), at different packing fractions are plotted in fig \ref{fig:etac_calculate}D for \(N = 25000, A = 5.0\). This calculation also exhibits a phase transition at \(\eta_c \approx 0.74 - 0.78\), as indicated by the change in the slope of the curve.

Therefore, we showed that the calculated values of positional ordering (RDF), defect counts, amplitude of Bragg peaks and the bond-orientational order parameter, using the pseudo-2D constrained positions of the centers of mass of the spherocylinders reveal similar values for the Liquid Crystal (LC) to \JDM{ordered} (or crystal(K)) transition packing fraction for the melting of a spherical monolayer of spherocylinders, with \(N = 25000\) and \(A = 5.0\). Hence, we take the average value of the upper and lower limits to be the transition packing fraction, with the difference between the two limits being the estimated error in our calculations. Hence for \(N=25000,A=5.0\), the transition occurs at \(\eta_c = 0.76 \pm 0.04\).

\subsection{Transition packing fraction and curvature strain}

\begin{figure}
\centering
\includegraphics[width=1.0\linewidth]{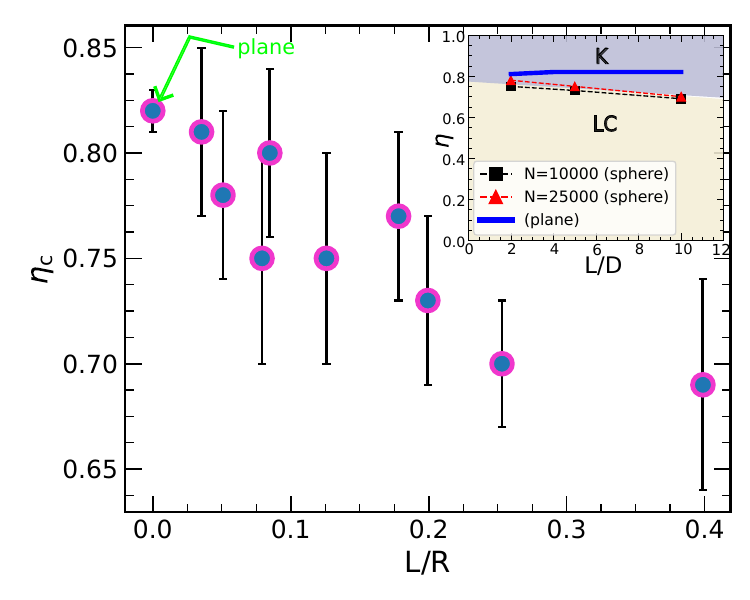}
\caption{Effect of curvature-induced strain on the transition packing fractions. As the strain, measured using the ratio \(L/R\), increases, \(\eta_c\) decreases. \JDM{Changing the system size or number of particles \(N\) changes the radius of the sphere, which gives rise to different \(L/R\) ratios for a specific aspect ratio of the spherocylinders (see table \ref{Tab:LbyR} first row for example). Similarly, changing the aspect ratios of the spherocylinders also changes the steric strain, as shown in table \ref{Tab:LbyR} in different rows. The plot shows the estimated \(\eta_c\) values for all such ratios.} Interestingly, for spheres which correspond to \(R >> L\),  with \(L/R \rightarrow 0\), the calculated value of \(\eta_c \) approaches \(\eta_c = 0.82\), which agrees with the planar monolayer results \cite{mandal2025freezing}, which is indicated by the data at \(L/R = 0\), as shown in the figure. The inset shows the phase diagram for the spherical monolayer of spherocylinders in the \((\eta, L/D)\) space. \JDM{In this figure, we showed the different phases, depending on the packing fraction and aspect ratios, for various system sizes \(N\).} The blue flat line in the diagram indicates the phase diagram for a flat monolayer of spherocylinders, which exhibit a constant transition packing fraction.}
\label{fig:transition_density_vs_LbyR}
\end{figure}

\begin{figure}
\centering
\includegraphics[width=0.95\linewidth]{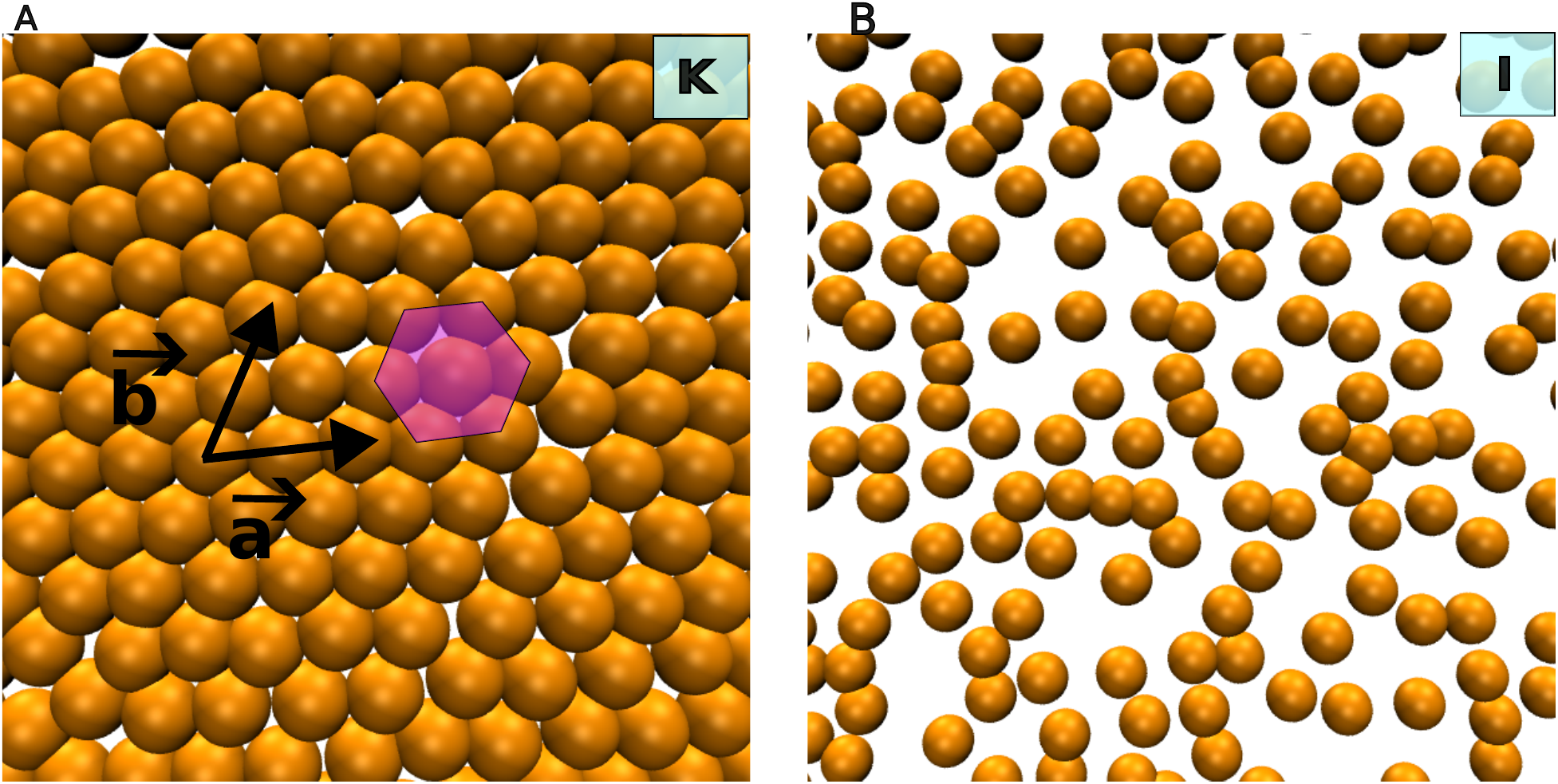}
\caption{Possibility of an intermediate hexatic phase: snapshots of the centers of mass for \(N=50000, L=2.0\). The 3-dimensional crystal phase of the spherical monolayer system corresponds to the 2d crystal (K) phase, whereas the 3D LC phase corresponds to a 2D isotropic (I) phase in the reference frame of the centers of mass of the spherocylinders. A) shows the in-plane positional hexagonal ordering of the 2D crystal (K) phase at \(\eta = 0.88\). For convenience, a pair of lattice vectors \(\Vec{a}\) and \(\Vec{b}\) are represented, along with a hexagon in purple color to demonstrate the underlying triangular lattice on the spherical surface. B) shows the disordered 2D isotropic (I) structure (\(\eta = 0.20\)) in the center of mass reference frame on the sphere. Hence, the transition, observed from the space of the centers of mass, can occur via an intermediate hexatic phase on the sphere. The snapshots show the view from very close to the top of the spherical surface and hence appear planar.}
\label{fig:hexatic_schematic}
\end{figure}

The values of \(\eta_c\) calculated above will depend on the values of an important dimensionless ratio of the confined system, namely \(L/R\).  This dimensionless ratio is a measure of the curvature-induced strain in the system and along with the length scale \(L/D\), this ratio also dictates the phase behaviors and transitions of the system. For example, if the radius of the confining sphere \(R\) is small, the curvature strain is large, and hence the steric strain at the tail-ends of the SRS particles in the interior of the sphere is larger. \JDM{Qualitatively, in order to reduce this strain, the system will prefer orientation defects \cite{rajendra2023packing} at even lower packing fractions which will give rise to the ordered phase.} This effect in turn reduces the transition packing fraction for the system. Thus, the values of \(\eta_c\) will depend on the values of \(L/R\) of the system. 

For our system, each melting simulation with a specific number of particles \(N\) and aspect ratio \(L/D\) gives rise to different values of the radius of the confining sphere \(R\), for each value of \(\eta\). But, in order to measure the representative curvature strain, we define the value of the ratio \(L/R\) for complete packing on the sphere \(\eta = 1.0\). Hence   
\begin{equation}
\eta =\frac{N}{16R^2} \\
     =1 \\
\implies L/R = 4L/\sqrt{N}  \label{eqn:LbyR} \\
\end{equation}

\JDM{Therefore, the steric strain depends both on the aspect ratio of the spherocylinders and the total number of particles.} In our simulations, we have studied the melting transition for three aspect ratios \(A = 2.0,5.0\) and \(10.0\), each for 3 different system sizes \(N = 10000,25000\) and \(50000\). For each of these case, the value of curvature-induced strain, measured by the ratio \(L/R\) for \(\eta=1.0\), is computed and is given in table \ref{Tab:LbyR}.


 
\begin{table}[]
    \centering
\begin{tabular}{ |c | c | c| } 
 \hline
 \textbf{\(A=L/D\)} & \textbf{\revv{\(N\)}} & \textbf{\revv{\(L/R\)}}   \\ \hline
  & 10000 & 0.08 \\ \hline
2 & 25000 & 0.0506 \\ \hline
 & 50000 & 0.0358\\ \hline
\hline

 & 10000 &  0.2\\ \hline
5 & 25000 & 0.1265 \\ \hline
 & 50000 & 0.0894\\ \hline
\hline
 
 & 10000 & 0.4 \\ \hline
10 & 25000 & 0.253 \\ \hline
 & 50000 & 0.1789 \\ \hline
    \end{tabular}
   
\caption{The measure of the values of the curvature-induced strain for each system and aspect ratio. As the number of particles in the system decreases, the radius of the confining sphere also reduces, which in turn increases the strain, as can be seen from the table.}
\label{Tab:LbyR}
\end{table}

For each of the systems mentioned in the table, we have computed the transition packing fraction using different quantities as mentioned in the previous subsection. In fig \ref{fig:transition_density_vs_LbyR}, we have \JDM{plotted the calculated values of \(\eta_c\) for each \(L/R\) ratio, which} shows the change in the transition packing fraction \(\eta_c\) with \(L/R\). We observe that as the curvature-induced strain increases in the system, the value of \(\eta_c\) decreases, as this helps in the removal of the steric clustering of the particles at the interior of the sphere. The reduction in the values of \(\eta_c\) seems to be linear with \(L/R\), although the exact nature of the plot is beyond the scope of the current study and thus was not studied further.

Interestingly, we also observe from fig \ref{fig:transition_density_vs_LbyR}, that as \(R >> L, L/R << 1\), the value of \(\eta_c\) approaches \(0.82\), which is the same as the transition packing fractions for planar monolayers \cite{mandal2025freezing}. Indeed, planar monolayers correspond to \(R \rightarrow \infty\), \(L/R \approx 0\), as indicated in fig \ref{fig:transition_density_vs_LbyR}. Remarkably, for such cases, the transition packing fraction is independent of the aspect ratio of the rods \JDM{which is demonstrated in the phase diagram in fig \ref{fig:transition_density_vs_LbyR} inset.} \JDM{Note that, to obtain the phase diagram, we fixed the total number of particles for a system and observed the different phases for a range of \(\eta\) values for different aspect ratios (\(L/D\)). For example, in fig \ref{fig:transition_density_vs_LbyR} inset, we show the phases for a spherical monolayer with \(N = 10000\) in red triangles, and similarly for \(N = 25000\) with black squares. The blue line shows the phases for the planar (or flat) monolayers, which is independent of \(N\).} The phase diagram (fig \ref{fig:transition_density_vs_LbyR} inset) for a spherical monolayer of spherocylinders demonstrates a trend similar to 3D bulk Isotropic-Nematic transition \cite{bolhuis1997tracing} and depends on the system size (due to the curvature strain), whereas the planar (or flat) monolayers show a transition packing fractions which is independent of system size or particle aspect ratio.

\subsection{Possibility of 2D hexatic phase near transition} \label{sec:hexatic_intro}

After computing the transition packing fractions for the systems, as mentioned above, we studied the nature of the phases for the system near the transition densities. Note that if we completely decouple the 3D orientational degrees of freedom of the spherocylinders and only consider the 2D degrees of freedom of the centers of mass on the spherical surface, the \JDM{"K"} phase indicates a (positionally) ordered distribution of the centers of mass (fig \ref{fig:hexatic_schematic}A) and the Liquid Crystalline phase is a (positionally) disordered isotropic phase (I)(fig \ref{fig:hexatic_schematic}B). Therefore, on the spherical surface, the 3D liquid crystal (LC) and crystal (K) phases  correspond to the 2D isotropic (I) and crystal (K) phases. Hence, the 3D K to LC transition, when observed from the space of the centers of mass, is similar to a crystal(K)-to-isotropic(I) transition on the curved surface. This observation gives rise to the question of whether an intermediate hexatic phase is possible for such systems near the transition packing fraction, where we might observe a sustained bond-orientational ordering without any positional ordering on the curved spherical surface.

To answer this question, we have used the following analyses. Note that, from now on, thecalculations are carried out on the center of mass positions of the particles, unless otherwise stated. Also, the "I" and "K" as shown in the subsequent figures are to be understood as the 2D isotropic and crystal phase which corresponds to the 3D LC and K phases respectively, for the 3D monolayer spherocylinder system, as mentioned earlier.

\subsection{2D Hexatic phase for spherical monolayer} In order to investigate the existence of the 2D hexatic phase for the spherical monolayer, we analyzed the structures for \(N=50000, A=2.0\), for which, \(\eta_c = 0.81 \pm 0.04\) (fig \ref{fig:transition_density_vs_LbyR}). \JDM{Some of the following analyses for the spherical monolayer were carried out at packing fractions close to \(\eta_c\), with a resolution of \(\Delta \eta = 0.02\).} 

\subsubsection{Order parameters:}

\begin{figure}
\centering
\includegraphics[width=1.0\linewidth]{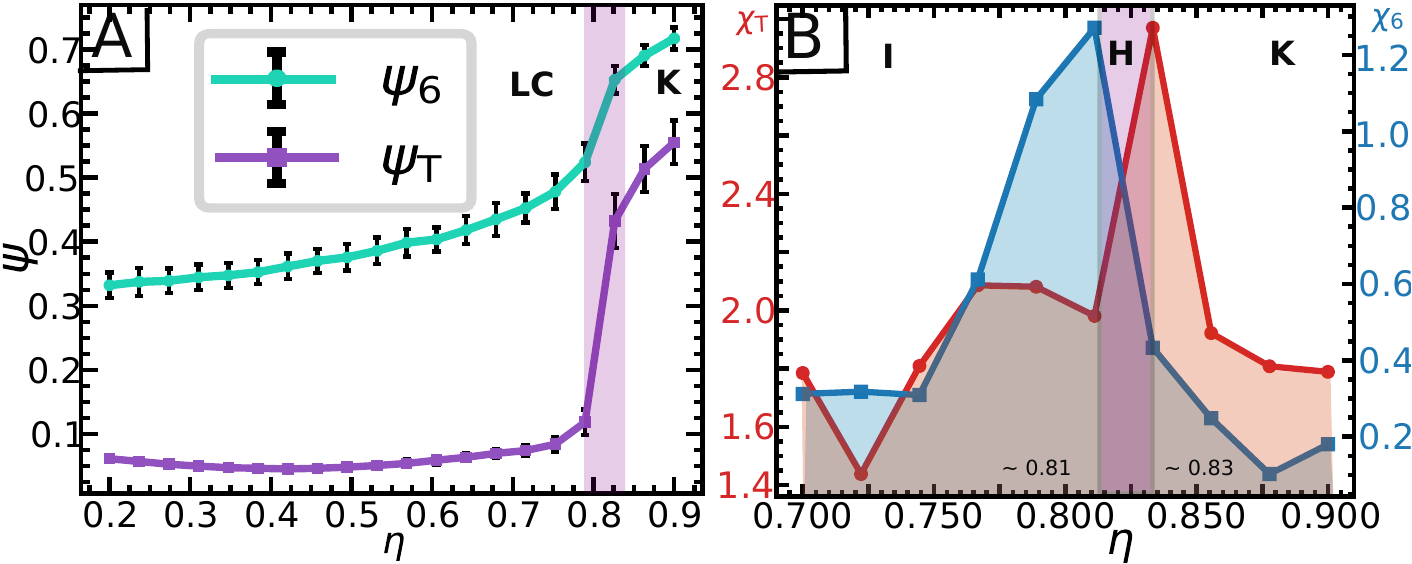}
\caption{Existence of a 2D hexatic phase for small rods on a spherical monolayer, \(L/D = 2.0, \eta_c = 0.81 \pm 0.04\): A) The translational (\(\psi_T\)) and bond orientational order parameter (\(\psi_6\)) as a function of packing fraction of the system. Both order parameters show a sharp rise at a packing fraction of \(\eta_c = 0.81\), indicated by a light purple rectangular strip in the plot. This indicates a 3D LC to K transition, which is similar to an I to K transition in 2D on the spherical surface. B) The corresponding susceptibilities, translational (\(\chi_T\), in red) and bond orientational (\(\chi_6\), in blue) are plotted with packing fraction for a range close to \(\eta_c\). \(\chi_6\) shows a peak at a packing fraction of \(\eta_6 = 0.81\) indicating a orientational ordering transition whereas \(\chi_T\) indicates an positional ordering transition in the structure at \(\eta_T = 0.83\), as is evident from the peak. This plot indicates the existence of a hexatic (H) phase at \(\eta = 0.81-0.83\), between the isotropic (I) (\(\eta<0.81\)) and crystal (K) (\(\eta>0.83\)) phases in 2D, as indicated in the plot.}
\label{fig:OP_suscep_N50k_L2}
\end{figure}

In order to quantify the ordering in the system, we have calculated both the bond orientational order parameter (\(\psi_6\)) (defined by eq \ref{eqn:psi6}) and the translational order parameter (\(\psi_T\)), which is defined as follows:

\begin{equation}
    \Psi_T = \frac{1}{N}\langle |\sum_k exp(i\Vec{G}\cdot\Vec{r_k})| \rangle  \label{eqn:psiT}\\
\end{equation}

where \(\Vec{G}\) is a first shell reciprocal lattice vector of the underlying triangular lattice and \(\Vec{r_k}\) corresponds to the position of the center of mass of the \(k-th\) molecule. We again emphasize that the bond-orientational (and also the translational ) order parameter we calculated is completely independent of the orientation degrees of freedom of the spherocylinders \(\Vec{s_i}\) and only takes into account the 2D structure of the centers of mass of the spherocylinders on the spherical surface.

In fig \ref{fig:OP_suscep_N50k_L2}A we plotted the translational (\(\psi_T\)) and bond orientational order parameters (\(\psi_6\), defined in eq \ref{eqn:psi6}) as functions of the packing fraction of the system for \(N = 50000, L/D = 2.0\). Both the order parameters show a phase transition at a packing fraction of around \(\eta_c \approx 0.81\). This transition corresponds to an LC to K transition in 3D, which is analogous to the I to K transition in 2D. We observe the hexatic order to be slightly higher than the translational order at all packing fractions. As noted earlier, the order parameters are calculated by dividing the spherical surface into smaller subsystems at different faces and then averaging over the results. \JDM{ We did not carry out calculations of correlation functions because we need larger face sizes to analyze the nature of the correlations at larger distances, but this makes it hard to define \(\phi_{jk}\) (see eq \ref{eqn:psi6}) or \(\Vec{G}\) (see eq \ref{eqn:psiT}) }

The corresponding susceptibility of the translational order parameter \(\chi_T\), which provides a definitive estimate of the translational ordering transition, is defined as follows:

\begin{equation}
    \chi_T = \frac{1}{N}\langle |\sum_k exp(i\Vec{G}\cdot\Vec{r_k})|^2 \rangle - N\Psi_T^2 
    \label{eqn:suscep2} \\
\end{equation}

To estimate the transition for the orientational order, we also computed the orientational susceptibility, \(\chi_6\), defined same as eqn \ref{eqn:suscep2}, with \(\psi_k\) replacing \(exp(i\Vec{G}\cdot \Vec{r_k})\), and \(\Psi_T\) replaced with \(\Psi_6\), where \(\psi_k\) and \(\Psi_6\) are defined in eqn \ref{eqn:psi6}.

In fig \ref{fig:OP_suscep_N50k_L2}B we show the corresponding susceptibilities, translational (\(\chi_T\)) and bond orientational (\(\chi_6\)) in the packing fraction range close to the value of \(\eta_c = 0.81\). \(\chi_6\) shows a peak at a packing fraction of \(\eta_6 = 0.81\) indicating an orientational disorder-order transition at this packing fraction. Hence, the system develops quasi-long-range orientational order at \(\eta > \eta_6\). On the other hand, \(\chi_T\) shows a peak at \(\eta = \eta_T = 0.83\), revealing a translational disorder-order transition at \(\eta_T\). At packing fraction below \(\eta_T\), positional order is short-ranged. Hence, at a packing fraction of \(\eta \in (0.81,0.83)\), there exists orientational order but no positional order. Fig \ref{fig:OP_suscep_N50k_L2}B) thus confirms the existence of an intermediate hexatic phase at a small packing fraction window of \(\Delta \eta = \eta_T - \eta_6 = 0.2\) in the melting transition of a spherical monolayer of spherocylinders in the reference frame of the sphere.

\begin{figure*}[tbhp]
\centering
\includegraphics[width=1.0\linewidth]{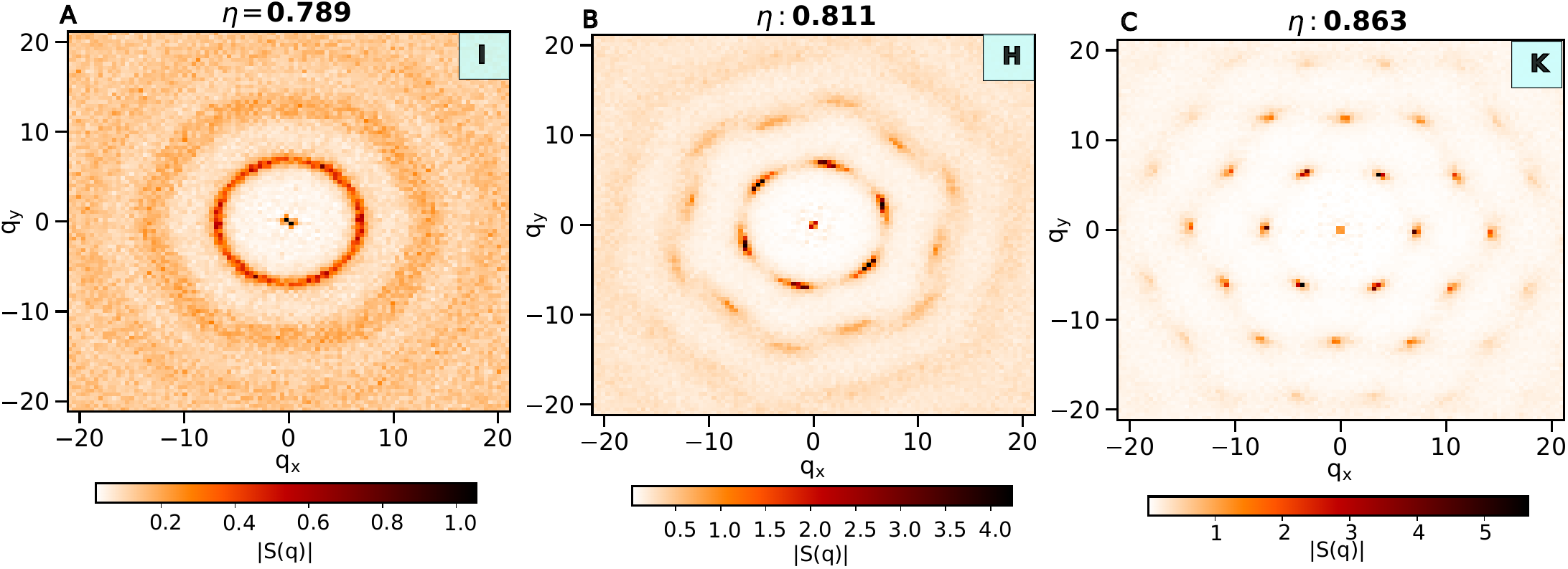}
\caption{Static structure factors for a spherical monolayer of spherocylinders at \(N=50000, L/D = 2.0\) at various packing fractions. A) The circularly symmetric distribution of the structure factor confirms an isotropic (I) phase at \(\eta = 0.789\).B) The diffuse 6-fold peaks in the structure factor at \(\eta = 0.811\) validates the existence of a hexatic phase. C) Sharp Bragg-like peaks start to appear at \(\eta = 0.826\), which indicates that a positional ordering transition occurs near this packing fraction.}
\label{fig:Str_factor_N50k_L2_T5}
\end{figure*}

\subsubsection{Structure factor}

The existence of the 2D hexatic phase in the packing fraction range of \(\eta \in (\eta_6, \eta_T)\), (\(\eta_6 = 0.81, \eta_T = 0.83\)) can also be confirmed from the static structure factor of the system, defined in eq \ref{eqn:str_fac}.


In fig \ref{fig:Str_factor_N50k_L2_T5} we have shown the static structure in the reciprocal space at three different packing fractions. Fig \ref{fig:Str_factor_N50k_L2_T5}A shows a circularly symmetric distribution of the structure factor in reciprocal space for \(\eta = 0.789\), which indicates a disordered or isotropic phase. In fig \ref{fig:Str_factor_N50k_L2_T5}B, we show that the structure factor shows diffused six-fold peaks at \(\eta = 0.811\), indicating the quasi-long-range orientational order and the six-fold symmetry of the hexatic phase. At a higher packing fraction of \(\eta > 0.826\), i.e. beyond the crystallization packing, sharper Bragg-like peaks start to appear in the system, indicating quasi-long-range translational order in two dimensions, which denotes a crystal phase.

\textit{Change in defect structures during melting.} We note that the melting of the 2D hexatic phase involves the disclination unbinding mechanism of the topological defects. For the system we studied, defects arise because of the frustration due to the spherical nature of the confining surface. Our analysis regarding whether some of these defects show disclination unbinding, is not conclusive. However, as is evident from fig \ref{fig:snapshot}B and fig \ref{fig:snapshot}D, there is a structural change in the defect arrangement. At high packing, the oppositely charged defects form dislocations, which attract each other to form domain boundaries, whereas at low packing fraction, the number of defects rise, and they are organised homogeneously over the surface of the sphere.

To summarize, we analyzed the hexatic and translational order parameters and the corresponding susceptibilities associated with the centers of mass of the spherocylinders, with an aspect ratio of \(A = 2.0\) on a sphere. These analyses show the presence of an intermediate 2D hexatic phase for the small rods on the surface of the sphere.

\subsection{2D Hexatic phase for a flat monolayer, L/R = 0}

After confirming the presence of the hexatic window in a spherical monolayer of spherocylinders, we now take the limiting case of \(R \rightarrow \infty\), which is the case of a flat monolayer.

\begin{figure*}
\centering
\includegraphics[width=1.0\linewidth]{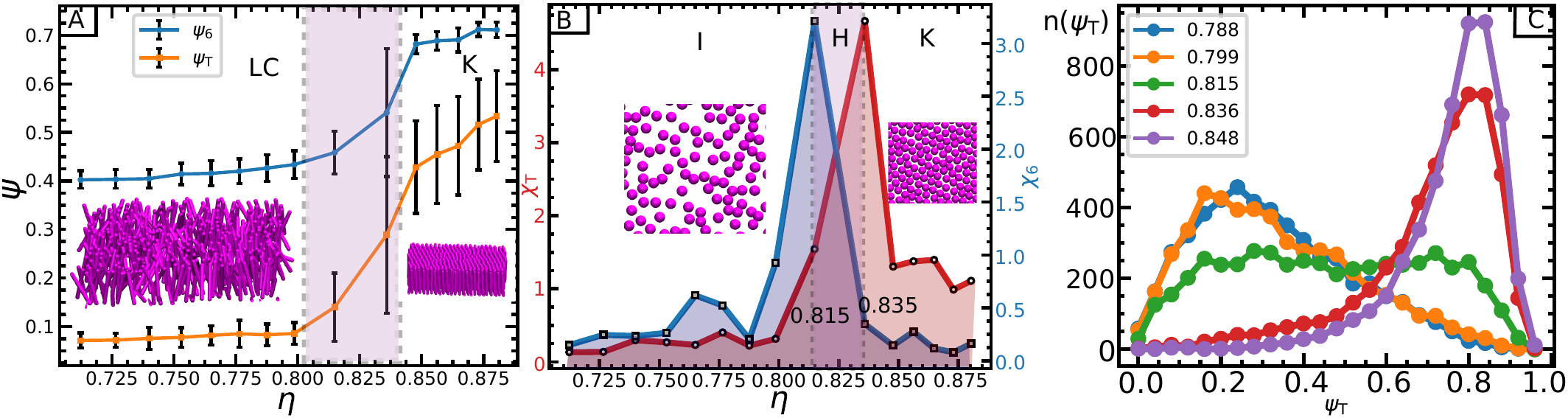}
\caption{Existence of hexatic phase for melting of a flat monolayer of spherocylinders, \(L/R = 0, A=L/D = 5.0\) from order parameter calculations. A) The translational and bond orientational order parameters as a functions of the packing fraction. The rise in the values of the order parameters indicate a positional and orientational ordering transition in the system at a packing fraction of \(\eta \approx 0.8 - 0.85\) which corresponds to a LC to crystal (K) phase transition in 3D, which in turn corresponds to an isotropic (I) to crystal (K) phase transition in 2D, in the center of mass reference frame on the flat surface. The corresponding structures of the 3D system for LC and K phases are also shown. B) The bond-orientational susceptibility \(\chi_6\) (in blue) and positional susceptibility \(\chi_T\) (in red) plotted against packing fraction. The peak in the plot of \(\chi_6\) and \(\chi_T\) indicating bond-orientational and positional transition of the system at \(\eta_6 = 0.815\) and \(\eta_T = 0.835\) respectively. Therefore an intermediate hexatic phase is present within the packing fraction range of \(\eta = 0.815-0.835\), indicated by the light pink shaded region. The packing fraction \(\eta < \eta_6, \eta \in (\eta_6,\eta_T) \) and \(\eta > \eta_T\) indicates  2D isotropic (I), hexatic (H) and crystal (K) phases respectively. The corresponding structures of the 2D isotropic (I) and crystal (K) phases are also shown. C) The distribution of the local translational order parameters at different packing fractions. The unimodality of the distribution indicates a continuous transition.}.
\label{fig:Flat_N25h_L5_T5_psi6_suscep_psiTdist}
\end{figure*}

\begin{figure*}
\centering
\includegraphics[width=1.0\linewidth]{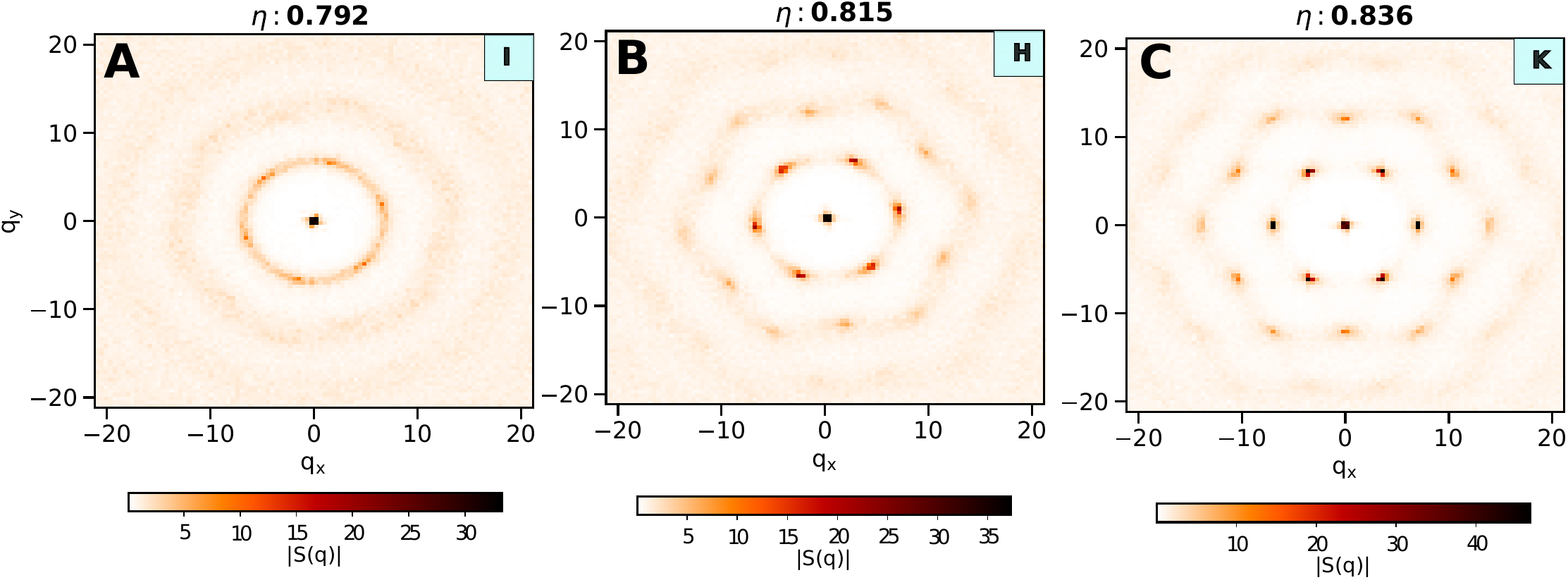}
\caption{Existence of 2D hexatic phase for melting of flat monolayer of spherocylinders (in the reference frame of the centers of mass), \(L/R = 0, A=L/D = 5.0\) from static structure factor calculations. A) Circularly symmetric distribution of the value of the structure factor \JDM{at the first Bragg shell} in reciprocal space indicating a disordered isotropic (I) phase at \(\eta = 0.732\). B) At \(\eta = 0.815\), the plot shows the emergence of diffuse hexagonally arranged peaks, while also retaining signatures of isotropic liquid nature. This indicates an intermediate hexatic phase (H). C) Sharp Bragg-like peaks appear at the specific reciprocal lattice vector values for \(\eta = 0.836\), a packing fraction in the crystal (K) domain.}.
\label{fig:Flat_N25h_L5_T5_str_fac}
\end{figure*}

\subsubsection{Order parameters:}

In fig \ref{fig:Flat_N25h_L5_T5_psi6_suscep_psiTdist}A we plot the variation of the order parameters with the packing fraction. The sharp rise in the values of the order parameters indicates a melting transition. Note that the bond-orientational order parameter and the translational order parameter, (and the corresponding susceptibilities discussed below) indicate an order-disorder transition in the 2D plane, in the center of mass reference frame of the spherocylinders, and the orientational degrees of freedom of spherocylinders in the third dimension is not considered in the calculation.

In order to determine the nature of the transition more accurately, in fig \ref{fig:Flat_N25h_L5_T5_psi6_suscep_psiTdist}B we show the corresponding susceptibilities, translational (\(\chi_T\)) and bond orientational (\(\chi_6\)) in the packing fraction range close to the value of \(\eta_c = 0.81\). Interestingly, both the plots show a sharp peak, but at two slightly different packing fractions. The hexatic susceptibility \(\chi_6\) shows a peak at a packing fraction \(\eta_6 = 0.815\), indicating a bond-orientational transition in the system. Similarly, the translational susceptibility also shows a peak, but at a slightly higher packing fraction of \(\eta_T = 0.835\). This indicates that the system undergoes a translational ordering transition at that packing fraction and configurations below \(\eta_T\) have no positional ordering. Therefore, the system at a packing fraction within the range \(\eta 
 = 0.815 - 0.835\) exhibits bond-orientational ordering but no translational ordering, indicating the presence of the hexatic phase at those packing fractions. It is interesting to note that the window of packing fraction in which the hexatic phase is sustained is \(\Delta \eta_H \approx 0.02\), which is similar to the values reported for the hexatic window for 2D soft WCA discs \cite{khali2021two}.

 We also calculated the translational order parameter \(\psi_T\) over subsystems spanning a small region on the surface of the confining plane and obtain their distribution function (unnormalised) \(n(\psi_T)\) (fig \ref{fig:Flat_N25h_L5_T5_psi6_suscep_psiTdist}C). The distribution function retains its unimodal form across the melting point \(\eta_T = 0.835\). As the solid melts, the peak of the distribution shifts to lower values. The unimodal nature of the distribution function indicates the absence of metastable phases beyond the transition, a defining characteristic of continuous transitions. At the hexatic melting transition at \(\eta_6 = 0.815\), the system shows a homogeneous distribution of the local translational order parameter.

\subsubsection{Structure factor}


We also computed the structure factor for the system for further analysis of the microscopic structure of the different phases. This computation provides further evidence for the presence of the hexatic phase for small rods.

In fig \ref{fig:Flat_N25h_L5_T5_str_fac}, we show the structure factors at three different packing fractions in three different phases of the system. Fig \ref{fig:Flat_N25h_L5_T5_str_fac} shows the nature of the structure factor for the isotropic packing fraction \(\eta = 0.792 < \eta_6 = 0.815\). The distribution shows a \JDM{circular} homogeneous pattern in the reciprocal space \JDM{at the first Bragg shell}, confirming the absence of positional and bond-orientational order in the system. At a packing fraction of \(\eta = 0.815\) (fig \ref{fig:Flat_N25h_L5_T5_str_fac}B), finite intensity peaks start to appear in the structure factor plot at six specific values of \((q_x,q_y)\). On the other hand, liquid-like homogeneous circular pattern is also readily observable in the plot for the first shell. This indicates that around the packing fraction of \(\eta = 0.815\), the system shows evidence of a hexatic phase. Characteristic of the hexatic phase are these six ring segments, a consequence of the quasi-long-range orientational order of sixfold symmetry. Beyond the crystallisation packing fraction of \(\eta_T = 0.835\), sharp Bragg-like peaks start to appear in the plot (fig \ref{fig:Flat_N25h_L5_T5_str_fac}C), mirroring the quasi-long-range character of translational order in two dimensions.

Therefore, to summarize, we find that for a flat monolayer of spherocylinders \((L/R \rightarrow 0, A = 5.0)\), when observed from the reference frame of the flat surface, there also exists an intermediate hexatic phase while the phase transition from an ordered crystal(K) to the disordered isotropic liquid (I) phase takes place. The solid melting occurs at \(\eta_T = 0.835\), whereas the hexatic melting occurs at \(\eta_6=0.815\).







\section*{Conclusion} \label{sec:conclusion}
In this work, we have worked with a system of soft repulsive spherocylinders (SRS particles, often mentioned as "rods" in the manuscript) with their centers of mass constrained on the surface of a sphere (and also a flat surface which corresponds to \(R \rightarrow \infty\), where \(R\) is the radius of the sphere). The system hence forms a spherical monolayer and the particles interact via the WCA (Weeks-Chandler-Anderson) interaction (\cite{weeks1971role}). First we observed that such systems exhibit two phases: Crystalline phase at high packing fraction, which is riddled with topological dislocations forming "domain scars", arising solely due to the topological confinement and a Liquid Crystalline phase which shows out-of-spherical-plane orientational order but no positional ordering in the centers of mass. The absence of 3-dimensional isotropic phase is due to the confinement effect.

Then we calculated the transition packing fractions using the analysis of: a) Radial Distribution function, b) total number of defect points, c) Hexatic order parameter and d) the amplitude of the static structure factor, and plotting them as functions of packing density. We showed that this transition depends on the system size due to the ratio \(L/R\), which is a measure of curvature-induced strain in the system. We showed that near to the lowest strain at large values of \(R\) (\(R>>L\)), the transition packing fraction approaches the value of that of planar monolayer (for which \(R \rightarrow \infty\)). Then we discussed the possibility of any intermediate hexatic phase in the system because when observed from the plane of the sphere, the centers of mass undergo a transition from ordered crystal to isotropic liquid while melting. Hence we carried out analyses in this regard.

Our analyses show that there exists an intermediate hexatic phase on the 2D surface for the system for both spherical and flat (\(R \rightarrow \infty\)) monolayer confinements.  For the case of a spherical monolayer with an aspect ratio of \(A=2.0\) of rods, calculations of hexatic and translational susceptibilities demonstrate the existence of an intermediate hexatic phase at \(\eta = 0.81-0.83\), which is confirmed by the structure factors. For a flat monolayer, we calculated the translational and hexatic order parameters and their corresponding fluctuations which indicate two different transition packing fractions: a 2D crystal to hexatic transition at \(\eta_T = 0.835\) and another transition from hexatic to 2D liquid at \(\eta_6 = 0.815\). The distribution of the local translational order parameter and the structure factors also confirmed the hexatic phase. Interestingly, 
We also showed that the defects that arise due to the spherical topology, also exhibit a change in structure during melting. At crystallisation packing, the oppositely charged defect pairs form a line, whereas at fluid packing, the defects are homogeneously distributed over the surface of the sphere.

Our research can be validated by studying phase transitions in various experimental systems, such as nanorods on different confining substrates. Further research work can also include, but are not limited to, a detailed study of the mechanism behind the emergence of the intermediate phases and also studying temperature-induced melting in such systems. Interesting melting scenarios might also occur for a system with high curvature strain as such high strains can alter the nature of the phase transition altogether, and thus they also present themselves as an interesting topic of future study. 





\section{Details of simulation} We have simulated a system of soft-repulsive spherocylinders (SRS, often termed "rods" in the manuscript), the center of masses of which are constrained to move on the surface of a sphere, using molecular dynamics simulations \cite{chattopadhyay2021heating,chattopadhyay2023two,chattopadhyay2024stability,chattopadhyay2023entropy}. A spherocylinder is a cylinder with hemispherical end-caps. The length and diameter of the cylinder are \(L\) and \(D\) respectively, and the aspect ratio of the rods is defined as \(A=L/D\). The interaction between the spherocylinders is taken to be of the WCA type \cite{weeks1971role}, given by the following potential 
\begin{align}
    U = \begin{cases}
            4\varepsilon\left[ \left( \frac{D}{d_m} \right)^{12} - \left( \frac{D}{d_m} \right)^6 \right] +\varepsilon, &d_m < 2^{1/6} D \\
            0 ,&d_m \geq 2^{1/6} D ,
        \end{cases}
\end{align}
where \(\epsilon\) is the energy scale and \(d_m\) is the distance of closest approach between any two spherocylinders \cite{vega1994fast}. The constraints on the center of masses of the particles are as follows:
\begin{align} 
    \left|  \Vec{r_i} \right| &= R, \label{eqn:rconstr} 
\end{align}
where \(\Vec{r_i}\) is the position of the center of mass of the \(i-th\) spherocylinder, and the origin of the co-ordinate system is taken to be at the center of the sphere of radius \(R\). For each system, simulations are done using the velocity Verlet  \cite{swope1982computer} and RATTLE \cite{andersen1983rattle} algorithms to solve the equation of motions of linear rigid molecules \cite{rapaport2004art}.

The simulations were carried out for three different system sizes \(N=10000,25000,50000\), where \(N\) is the total number of particles in the system. For each of these systems, we simulated for three different aspect ratios of the rods, \(A=2.0,5.0,10.0\). The simulations started with high packing fraction, \(\eta=0.9\), where \(\eta = N/16R^2\), in an NVT ensemble. The simulations were run for \(0.2 \times 10^6\) steps and the last half of these steps were used to do the analyses. After equilibrating the system at a specific packing fraction, its final structure was used for the next NVT run for the lower packing fraction, in order to reduce the equilibration time. Using this technique, we simulated the pressure-induced melting in the system for all the cases.

We have used reduced units for our calculations throughout the work. The temperature are defined as follows: \(T^* = k_BT/\epsilon\), where \(k_B\) is the Boltzmann constant, \(T\) is the temperature of the system obtained from the equipartion theorem and \(\epsilon\) is as defined above. For simplicity, we have used the values \(\epsilon=1.0,D=1.0\). The simulations were carried out at a constant reduced temperature \(T^*=5.0\).

\subsection*{Data Availability}
 The data that support the findings of this study are available from the corresponding author upon reasonable request.

\begin{acknowledgments}

JM thanks MHRD, India for the fellowship. PKM thanks DST, India for financial support and SERB, India for funding and computational support. CD thanks SERB, India for a SERB Distinguished Fellowship.
\end{acknowledgments}

\section*{Conflict of Interest}
The authors have no conflicts to declare.
\section*{Author Contribution}
J.M. and P.K.M. designed and conceptualised research; J.M. performed research and analyzed data; C.D. contributed to discussions and provided feedback on the analysis; P.K.M. provided resources; P.K.M. and C.D. supervised research; and J.M., C.D., and P.K.M. wrote the paper.



\bibliography{ref}


\appendix

\renewcommand\thefigure{\thesection.\arabic{figure}}    
\
%
\end{document}


\newgeometry{top=1.50 cm,bottom=1.50 cm,left=1.50 cm,right=1.25 cm}

\maketitle

\section{Compressibility reduction for charged rods}

As expected, the compressibility is reduced for a system of charged rods for small electrostatic interaction amplitudes that we explored ($u<0.1$). This is shown in \fig{fig:EOS_Lall_uall_kall} below. The effect is most clearly reflected for long rods $A=10$.

\begin{figure}[h!]
    \centering
    \includegraphics[width=0.6\linewidth]{Figures/EOS_Lall_uall_kall.pdf}
    \caption{ Overview of the equation of state in terms of reduced preesure $P^{\ast}$ versus packing fraction $\eta$ for a rod monolayer for both neutral and charged with various aspect ratios $A$. The electrostatic amplitude is given by $u$ while $\kappa$ denotes the Debye screening length in units of the rod diameter. }
    \label{fig:EOS_Lall_uall_kall}
\end{figure}



\section{Conically degenerate tilt angle for charged rods}

The orientational distribution and the tilt angle for  charged rods  show qualitatively similar behaviour to the neutral case ($u=0$). For the weak amplitudes considered ($u\leq 0.1$) the electrostatic interaction can be considered as a perturbation for the considered set of values of \(u\). In \fig{fig:OD_compare_tilt_angle_el_on_off_compare} we demonstrate that the tilt angle \(\theta_t\) decreases with the electrostatic interaction strength \(u\) for a specific value of \(A\) and \(\eta\).

\begin{figure}[h!]
    \centering
    \includegraphics[width=0.6\linewidth]{Figures/OD_compare_tilt_angle_el_on_off_compare.pdf}
    \caption{(A)-(C) Effect of twist electrostatic repulsion on the rod orientational probability for a variety of rod aspect ratios \(A=2.0,5.0\) and \(10.0\), electrostatic amplitudes ranging from neutral to weakly charged rods (\(u=0.0,0.01,0.1\)) and Debye screening lengths $\kappa$. (D) Variation of the tilt angle  with packing fraction $\eta$.  }
    \label{fig:OD_compare_tilt_angle_el_on_off_compare}
\end{figure}